\documentclass[]{pasj01}       
\usepackage{bm}
\newcommand{\sisuu}[1]{\times 10^{#1}}
\newcommand{\average}[1]{\ensuremath{\langle#1\rangle} }
\begin{document}        
\Received{$\langle$reception date$\rangle$}
\Accepted{$\langle${2018/Sep/03}$\rangle$}
       
\title{Systematic two-dimensional radiation hydrodynamic simulations of super-Eddington accretion flow and outflow: Comparison with the slim disk model
  }

\author{Takaaki \textsc{Kitaki},\altaffilmark{1}$^{,*}$       
  Shin \textsc{Mineshige},\altaffilmark{1}       
  Ken \textsc{Ohsuga},\altaffilmark{2}       
  and Tomohisa \textsc{Kawashima}\altaffilmark{3}       
}       
  \altaffiltext{1}{Department of Astronomy, Graduate School of Science, Kyoto University, Kitashirakawa-Oiwake-cho, Sakyo-ku, Kyoto 606-8502, Japan}       
  \altaffiltext{2}{Center for Computational Sciences, University of Tsukuba, Ten-nodai, 1-1-1 Tsukuba, Ibaraki 305-8577, Japan}
  \altaffiltext{3}{National Astronomical Observatory of Japan, 2-21-1 Osawa, Mitaka-shi, Tokyo 181-8588, Japan}         
  \email{kitaki@kusastro.kyoto-u.ac.jp}       
       
       
       
\maketitle       
       
\begin{abstract}
To what extent can the one-dimensional slim disk model reproduce the multi-dimensional results of global radiation-hydrodynamic simulations of super-Eddington accretion?
With this question in mind, we perform systematic simulation study of accretion flow onto a non-spinning 
black hole for a variety of black hole masses of $(10 - 10^7) M_{\odot}$
and mass accretion rates of $(1.4 \times 10^2 - 5.6 \times 10^3)L_{\rm Edd}/c^2$ 
(with $L_{\rm Edd}$ and $c$ being the Eddington luminosity and speed of light).
In order to adequately resolve large-scale outflow structure,
we extensively expand a simulation box 
to cover the space of $3000 r_{\rm S}$ (with $r_{\rm S}$ being the Schwarzschild radius), 
larger than those in most previous studies, so that we can put relatively large angular momentum
to the gas injected from the outer simulation boundary.
The adopted Keplerian radius, at which the centrifugal force balances with the gravitational force,
is $r_{\rm K}=300 r_{\rm S}$. The injected mass first falls and is accumulated at around 
this radius and then slowly accretes towards the central black hole via viscosity.
We simulate such accretion processes, taking inverse and bulk Compton scattering into account.
The simulated accretion flow is in a quasi-steady state inside $r_{\rm qss}\sim 200 r_{\rm S}$. 
Within this radius the flow properties are, as a whole, in good agreement with those 
described by the slim disk model
except that the radial density profile of the underlying disk is much flatter, $\rho\propto r^{-0.73}$ 
(cf. $\rho\propto r^{-3/2}$ in the slim disk model), due probably to efficient convection.
We find very weak outflow from inside $r\sim 200 r_{\rm S}$ unlike the past studies.
\end{abstract}      
            
\section{Introduction}
It has been long believed that the Eddington luminosity is a classical limit to the luminosities of 
any accreting objects. However, we now know that this is no longer the case
both from the observational and theoretical grounds.
In fact, there are growing observational evidences supporting the existence of super-Eddington
accretion in several distinct classes of objects. 
In parallel with extensive observational studies multi-dimensional simulations are
being performed by a number of groups. 
Super-Eddington (or supercritical) accretion is attracting much attention among researchers.

Good candidates for super-Eddington accretors are ultra-luminous X-ray sources (ULXs),
bright X-ray compact sources of X-ray luminosity of $10^{39}-10^{41}[{\rm erg ~s^{-1}}]$
(see Kaaret, Feng, \& Roberts 2017 for a recent review and references therein).
Quite a few ULXs have been discovered so far in off-nuclear regions of nearby galaxies
and its number is rapidly increasing.
There are two main routes of idea to explain their high luminosities:
the sub-Eddington accretion onto the intermediate-mass black holes (Makishima et al. 2000, Miller et al. 2004),
and the super-Eddington accretion onto the stellar mass black holes (Watarai et al. 2001, King et al. 2001).
(Both scenarios require high accretion rates, but this is a separate issue and we do not go into details.)
The discovery of ULX pulsars which the central object is the neutron star supports the latter scenario
(NGC7793 P13, F\"{u}rst et al. 2016, Israel et al. 2017b; NGC5907 ULX, Israel et al. 2017a; NGC300 ULX-1, Kosec et al. 2018).

Other super-Eddington accretors are found in narrow-line Seyfert 1 galaxies (NLS1s),
bright micro-quasars such as GRS1915+105, ultra-soft X-ray sources (ULSs), etc.
The NLS1s harbor less massive central black holes
(with mass $\lesssim 10^{7}M_{\odot}$, see Boller et al. 2000) 
than broad-line Seyfert 1 galaxies (BLS1s).
That is, the NLS1s tend to have higher Eddington ratios than the BLS1s with similar luminosities, 
thus super-Eddington accretion being more feasible in the former
(Wang \& Zhou 1999; Mineshige et al. 2000).
Jin et al. (2017), for example, analyzed RX J0439.6-5311 (NLS1) using the multi-wavelength spectrum
and estimated the accretion rate to be $\sim 71 L_{\rm Edd}/c^{-2}$ in the outer disk
(for the black hole mass $\sim 1\times 10^{7}M_{\odot}$).
The black hole binary, GRS1915+105, is also known to stay occasionally in the super-Eddington phase
(see, e.g., Done et al. 2007; Vierdayanti et al. 2010).

One of the most prominent features of the super-Eddington flow is photon trapping
(Katz 1977; Begelman 1978).
That is, photons generated in the accretion disk
tends to be directly swallowed by a black hole
before diffusing toward the surface of the disk.
This occurs, when accretion rate is high, since then
the diffusion time $t_{\rm diff}$ becomes longer than the accretion time $t_{\rm acc}$.
The equation $t_{\rm diff}=t_{\rm acc}$ leads to photon trapping radius $r_{\rm trap}$ (e.g. Ohsuga et al. 2002),
\begin{equation}
r_{\rm trap}\equiv\frac{3}{2} \frac{\dot{M}}{L_{\rm Edd}/c^{2}} \frac{H}{r}r_{\rm S}. \label{eq-r-trap}
\end{equation}
Here, $\dot{M}$ is the mass accretion rate,
and $H$ is the half thickness of the accretion disk.
The diffused photons in the region $r<r_{\rm trap}$ accrete into black hole with gases 
because photons cannot escape from the accretion disk.
We wish to stress here that photon trapping is essentially multi-dimensional effect
(Ohsuga et al. 2002). This is a key issue to be discussed in the present paper.

The slim disk model is the one-dimensional accretion disk model
including photon-trapping effect as the advection of the photon entropy in the energy equation
(Abramowicz et al. 1988; see Chap. 10 of Kato et al. 2008 for a concise review).
The basic equations for the radial structure of the slim disk are derived from 
the Navier--Stokes equations under the vertically one-zone approximation;
that is, physical quantities are integrated in the $z$-direction (perpendicular to the equatorial plane).
The slim disk model is a numerical model, but approximate analytical expression 
is available, which was obtained by Watarai (2006).
They gave simple formula describing the parameter dependence of mass density, temperature, velocity
on the equatorial plane of the slim disk model.

Outflow is another signature characterizing super-Eddington flow 
(see a pioneering discussion by Shakura \& Sunyaev 1973).
When the disk luminosity exceeds the Eddington luminosity $L_{\rm Edd}$,
this means that the radiation force is greater than gravity by the definition of the Eddington luminosity.
A part of the accretion flow gas is blown out from the disk accelerated by radiation pressure,
then the outflow occurs in high mass accretion rate.
Note that such multi-dimensional motion as large-scale circulation (convection)
and outflow are not explicitly considered in the slim disk model.
We thus need to perform multi-dimensional radiation hydrodynamic (RHD) simulations.

Ohsuga et al. (2005) pioneered the two-dimensional RHD simulations for super-Eddington accretion flow
(see also Kawashima et al. 2009, 2012, Hashizume et al. 2015, Ogawa et al. 2017, Kitaki et al. 2017 for RHD simulations;
Ohsuga et al. 2009, Ohsuga \& Mineshige. 2011, Jiang et al. 2014 for radiation-magneto hydrodynamic (R-MHD) simulations;
S\c{a}dowski et al. 2014, 2015, McKinney et al. 2014, Fragile et al. 2014, Takahashi et al. 2016
for general relativistic radiation-magneto hydrodynamic (GR-R-MHD) simulations). 
In these simulation studies the authors adopted somewhat unrealistic situations;
that is, they commonly assume relatively small angular momentum of accreting gas, 
which is either injected from the outer simulation boundary or provided from the initial gaseous torus.
This was necessary for numerical reason, since otherwise it will take enormous computation time
(corresponding to a long viscous timescale) to be completed within a few months.
This leads to a quite narrow viscous accretion region in a quasi-steady state 
(within a few tens of Schwarzschild radii in most cases), 
which makes it difficult to compare with slim disk model

In the present study, therefore,
the initial angular momentum is set larger.
We calculate the structure of the super-Eddington accretion flow and outflow 
by means of the two-dimensional (2D) radiation hydrodynamic (RHD) simulations
for a variety of black hole masses ($M_{\rm BH}$) and mass accretion rates ($\dot M_{\rm BH}$), 
and compare the simulation results with those calculated based on the slim disk model.
The main feature in this study compared with previous papers is
the systematic study of the scaling relations produced by numerical simulations,
which is due to the volume of parameter space spanned (along with a larger radial extension) more than
to a higher spatial resolution (some of the studies presented in the introduction actually use better resolution).
We also obtained the fitting formulas of the super-Eddington accretion disk for the first time.

The plan of this paper is as follows: We first explain our models and methods of calculations
in the next section. We then show our main results in section 3 and discussion in section 4.

\section{Models and Numerical Methods}
\subsection{Radiation Hydrodynamic Simulations}
In the present study, we consider super-Eddington accretion flow and outflow onto a black hole 
by injecting mass from the outer simulation boundary at a constant rate of $\dot{M}_{\rm input}$
with a certain amount of angular momentum.
(The parameter values will be specified in section 2.2.)
The flux-limited diffusion approximation is adopted (Lervermore \& Pormaraning 1981; Turner \& Stone 2001).
We also adopt the $\alpha$ viscosity prescription (Shakura \& Sunyaev 1973) 
and assign $\alpha=0.1$ throughout the present study.
General relativistic effects are incorporated by adopting the pseudo-Newtonian potential (Paczy\'{n}sky \& Wiita 1980).

Basic equations and numerical methods are the same as those in Kawashima et al. (2009, 2012),
but it is upgraded to solve energy equations with implicit method.
This 2D-RHD code solves the axisymmetric two-dimensional radiation hydrodynamic equations     
in the spherical coordinates $(x,y,z)=(r\sin\theta\cos\phi, r\sin\theta\sin\phi, r\cos\theta)$,
where the aziumuthal angle $\phi$ is set to be constant.
The continuity equation is given by,
\begin{eqnarray}
  \frac{\partial \rho}{\partial t}+\nabla\cdot\left(\rho\bm{v}\right)=0.
\end{eqnarray}
Here, $\rho$ is the gass mass density,
$\bm{v}=(v_{r},v_{\theta},v_{\phi})$ is the gas velocity. 
The equations of motion are written as,
\begin{eqnarray}
  \frac{\partial (\rho v_{r})}{\partial t}+\nabla\cdot\left(\rho v_{r}\bm{v}\right)=-\frac{\partial p}{\partial r}+\rho\left(\frac{v_{\theta}^{2}}{r}+\frac{v_{\phi}^{2}}{r}-\frac{GM_{\rm BH}}{(r-r_{\rm s})^{2}}\right)\nonumber\\
  +\frac{\chi}{c}F_{0r},\\
  \frac{\partial (\rho rv_{\theta})}{\partial t}+\nabla\cdot\left(\rho rv_{\theta}\bm{v}\right)=-\frac{\partial p}{\partial \theta}+\rho v_{\phi}^{2}\cot\theta\nonumber\\
  +r\frac{\chi}{c}F_{0\theta},\\
  \frac{\partial (\rho r\sin\theta v_{\phi})}{\partial t}+\nabla\cdot\left(\rho r\sin\theta v_{\phi}\bm{v}\right)=\frac{1}{r^{2}}\frac{\partial}{\partial r}\left(r^{3}\sin\theta t_{r\phi}\right),
\end{eqnarray}
$p$ is the gass pressure,
$r_{\rm s}\equiv 2GM_{\rm BH}/c^{2}$ is the Schwarzschild radius
with $G$ being the gravitational constant and $c$ being the light speed.
$\chi=\kappa+\rho \sigma_{\rm T}/m_{\rm p}$ is the total opacity,
where $\kappa$ is free-free and free-bound absorption opacity
(Rybicki \& Lightman 1979),
$\sigma_{\rm T}$ is the cross-section of Thomson scattering,
and $m_{\rm p}$ is the proton mass.
$\bm{F}_{0}=(F_{0r},F_{0\theta})$ is the radiative flux in the comoving frame,
where the suffix 0 represents quantities in the comoving frame.
Using the dynamical viscous coefficient $\eta$,
$t_{\rm r\phi}$ is the viscous stress tensor described as
\begin{eqnarray}
  t_{r\phi}&=&\eta r \frac{\partial }{\partial r}\left(\frac{v_{\phi}}{r} \right),\\
  \eta&=&\alpha \frac{p+\lambda E_{0}}{\Omega_{\rm K}}.
\end{eqnarray}
Here, $\Omega_{\rm K}$ is the Keplerian angular speed,
$E_{0}$ is the radiation energy density,
and $\lambda$ represents the flux limiter of the flux-limited diffusion approximation
(Levermore \& Pormraning 1981; Turner \& Stone 2001).

The energy equations of the gas and the radiation are given by,
\begin{eqnarray}
  \frac{\partial e}{\partial t}+\nabla\cdot\left(e\bm{v}\right)&=&-p\nabla\cdot\bm{v}-4\pi\kappa B+c\kappa E_{0}\nonumber\\
  &&+\Phi_{\rm vis}-\Gamma_{\rm Comp},\\
  \frac{\partial E_{0}}{\partial t}+\nabla\cdot\left(E_{0}\bm{v}\right)&=&-\nabla\cdot\bm{F}_{0}-\nabla\bm{v}:\bm{P}_{0}+4\pi\kappa B-c\kappa E_{0}\nonumber\\
  &&+\Gamma_{\rm Comp}.
\end{eqnarray}
Here, $e$ is the internal energy density
which is linked to the thermal pressure by the ideal gas equation of state,
$p=(\gamma -1)e=\rho k_{\rm B}T_{\rm gas}/(\mu m_{\rm p})$
with $\gamma=5/3$ being the specific heat ratio,
$k_{\rm B}$ being the Boltzmann constant,
$\mu=0.5$ is the mean molecular weight,
and $T_{\rm gas}$ is the gas temperature.
$B=\sigma_{\rm SB}T_{\rm gas}^{4}/\pi$ is the blackbody intensity
where $\sigma_{\rm SB}$ is the Stefan--Boltzmann constant.
$\bm{P}_{0}$ is the radiation pressure tensor,
$\Phi_{\rm vis}$ is the viscous dissipative function written as
\begin{eqnarray}
  \Phi_{\rm vis}=\eta\left[r\frac{\partial }{\partial r}\left(\frac{v_{\phi}}{r} \right)\right]^{2}
\end{eqnarray}
The Compton cooling/heating rate $\Gamma_{\rm Comp}$ is described as
\begin{eqnarray}
  \Gamma_{\rm Comp}&=&4\sigma_{\rm T}c\frac{k_{\rm B}\left(T_{\rm gas}-T_{\rm rad}\right)}{m_{\rm e}c^{2}}\left(\frac{\rho}{m_{\rm p}} \right)E_{0}
\end{eqnarray}
Here, $m_{\rm e}$ is the electron mass and
$T_{\rm rad}\equiv (E_{0}/a)^{1/4}$ is the radiation temperature with the radiation constant $a=4\sigma_{\rm SB}/c$.

Simulation settings are roughly the same as those in Kitaki et al. (2017)
except fot the initial conditions and the size of the simulation box.
The computational box is set by $r_{\rm in}=2r_{\rm S}\leq r \leq r_{\rm out}=3000r_{\rm S}$,    
and $0 \leq \theta \leq\pi/2$.
Grid points are uniformly distributed in logarithm in the radial direction;
$\triangle\log_{10} r = (\log_{10} r_{\rm out}-\log_{10} r_{\rm in})/N_{r}$,
while it is uniformly distributed in $\cos\theta$ in the azimuthal direction;
$\triangle\cos \theta=1/N_{\theta}$, where the numbers of grid points     
are $(N_{r},N_{\theta})=(192,192)$ throughout the present study.    
    
\subsection{Initial conditions and calculated models}
\label{sec-initial-condition}
All the parameter values of calculated models are summarized in Table \ref{table1}.
We start calculations with an empty space around a black hole with mass of $M_{\rm BH}$,
though we initially put a hot optically thin atmosphere with negligible mass for numerical reasons.
Mass is injected continuously with a constant rate of $\dot{M}_{\rm input}$
through the outer disk boundary at $r=r_{\rm out}$ and $0.45\pi\leq\theta\leq0.5\pi$.  
The black hole mass and mass injection rate are free parameters and are set to be
$M_{\rm BH}=10$, $10^4$, and $10^7$ ($M_{\odot}$), 
and $\dot{M}_{\rm input} = 3\times 10^2$, $10^3$, $5\times10^{3}$, $10^4$, and $10^5$ ($L_{\rm Edd}/c^2)$.
We set the injected mass to have an angular momentum corresponding to the 
Keplerian angular momentum at the Keplerian radius, $r=r_{\rm K}$, which is a free parameter; 
that is, the initial specific angular momentum is $\sqrt{GM_{\rm BH} r_{\rm K}}$.
We thus expect that inflow material first falls towards the center
and forms a rotating gaseous ring at around $r\sim r_{\rm K}$, 
from which the material slowly accretes inward via viscous diffusion process.
We allow mass to go out freely through the outer boundary at $r=r_{\rm out}$
and $0\leq\theta\leq0.45\pi$.
and assume that mass at $r=r_{\rm in}$ is absorbed.

\begin{table}[h]
  \tbl{Calculated Models}{    
    \begin{tabular}{cccc}
      \hline     
      model & $M_{\rm BH}$ & $\dot{M}_{\rm input}$ & $r_{\rm K}$  \\
      name & [$M_{\odot}$] & [$L_{\rm Edd}/c^{2}$] & [$r_{\rm S}$]\\
      \hline
      (a11) &  & $300$ & $100$ \\
      (a12) &  & $10^{3}$ & $300$\\
      (a13) & $10^{1}$ & $5\times 10^{3}$ & $300$\\      
      (a14) &  & $10^{4}$ & $300$\\
      (a15) &  & $10^{5}$ & $300$\\\hline
      (a41) &  & $300$ & $100$ \\
      (a42) &  & $10^{3}$ & $300$\\
      (a43) & $10^{4}$ & $5\times 10^{3}$ & $300$\\
      (a44) &  & $10^{4}$ & $300$\\
      (a45) &  & $10^{5}$ & $300$\\\hline
      (a71) &  & $300$ & $100$  \\
      (a72) &  & $10^{3}$ & $300$\\
      (a73) & $10^{7}$ & $5\times 10^{3}$ & $300$\\      
      (a74) &  & $10^{4}$ & $300$\\
      (a75) &  & $10^{5}$ & $300$\\
      \hline
  \end{tabular}}
  \begin{tabnote}    
    Here, $M_{\rm BH}$ is the black hole mass,
    $\dot{M}_{\rm input}$ is the mass injection rate,
    $r_{\rm K}$ is the Kepler radius (see text).
    The other parameters are same among all models; 
    the inner radius is $r_{\rm in}=2r_{\rm S}$,
    the outer radius is $r_{\rm out}=3000r_{\rm S}$,
    the $\alpha$ viscosity is $\alpha=0.1$, and the metallicity is $Z=Z_{\odot}$.
  \end{tabnote}
  \label{table1}     
\end{table}

\section{Results}   
\subsection{Overall flow structure}       
\begin{figure*}[p]
\begin{center}
  \includegraphics[width=160mm,bb=0 0 2834 1700]{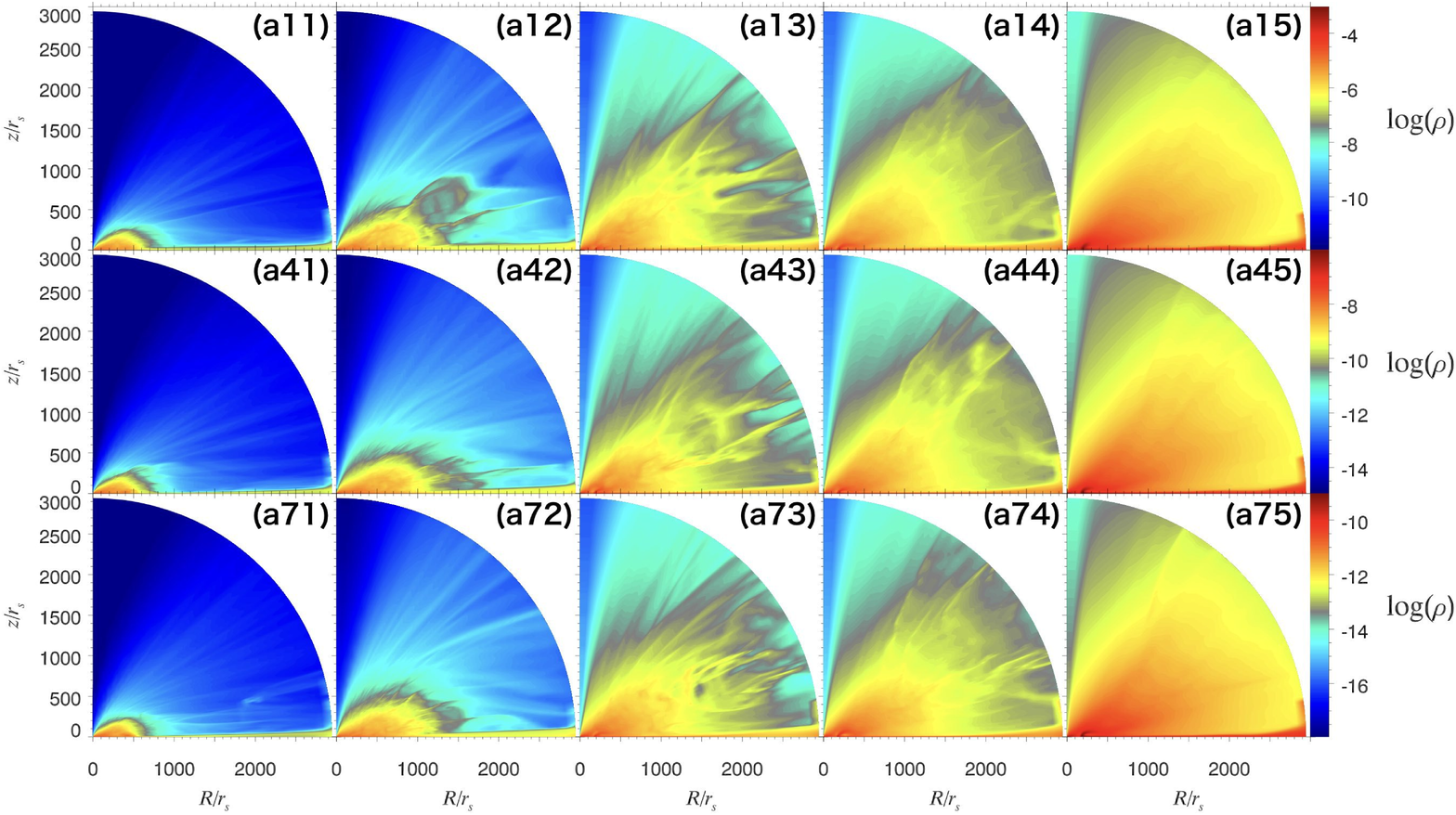}
\end{center}
\caption{Time-averaged density contours of super-Eddington accretion flow onto black holes
  with masses of $M_{\rm BH} = 10^{1}, 10^{4},$ and $10^{7} (M_{\odot})$ from the top to the bottom,
  for the mass injection rates of $\dot{M}_{\rm input} = 300, 10^{3}, 5\times10^{3}, 10^{4}, 10^{5} (L_{\rm Edd}/c^{2})$
  from the left to the right, respectively.
  Note different color scales for different black hole masses.
  Each panels looks similar, if we adjust the color scale according to the density normalization
  relationship of $\rho_0\propto M_{\rm BH}^{-1}$ (see Kitaki et al. 2017).
}
\label{fig1}
\begin{center}
  \includegraphics[width=160mm,bb=0 0 2834 1700]{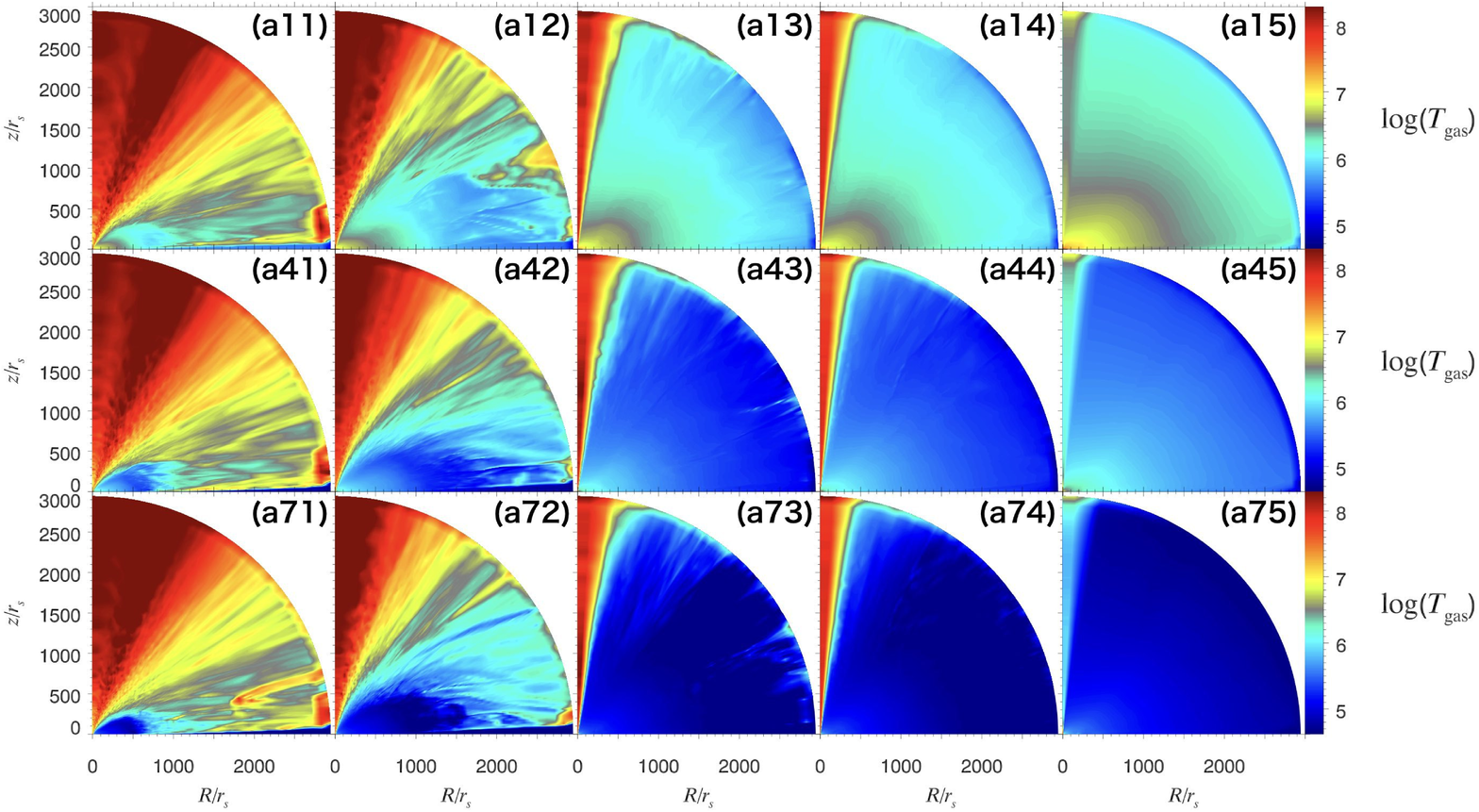}
\end{center}
\caption{Same as Figure \ref{fig1} but for the time-averaged gas temperature distributions.
  The color scales are the same in all the panels.
}
\label{fig2}
\end{figure*}

We first overview the simulated flow structure plotted in Figures \ref{fig1} and \ref{fig2}.
These are the two-dimensional color contours of the time averaged mass density and gas temperature
of all the calculated models, respectively.
As is clearly seen in Figure \ref{fig1}, the injected gas accumulates at around $r_{\rm K}$
($\sim 300 r_{\rm S}$ except for the panels at the left end)
and forms a puffed up structure inside $\sim 10^3 r_{\rm S}$. 
The accumulated gas then slowly accretes towards the center.
After a sufficient time over the viscous timescale,
an accretion flow settles in a quasi-steady state (Ohsuga et al. 2005).

If we see the panels in Figure \ref{fig1} from the left to the right,
we notice that the puffed up region grows with the increase of the mass injection rate,
although the Keplerian radius is kept the same (except in the panels at the left end).
This is because the trapping radius increases with an increase in the mass accretion rate,
obeying $r_{\rm trap} \sim (\dot{M}_{\rm BH}c^2/L_{\rm Edd})r_{\rm S}$.
If we see the panels from the top to the bottom in Figure \ref{fig1}, on the other hand,
we see no big changes, as long as we change the color scale according to the relationship,
$\rho_0 \propto M_{\rm BH}^{-1}$.
This scaling of density with the blackhole mass is linked to mass conservation (Kitaki et al. 2017).

Let us next compare different models in terms of the temperature distribution displayed 
in Figure \ref{fig2}.
As we see from the left to the right panels, we understand that the high temperature regions
(with red color) shrink as $\dot M_{\rm BH}$ increases, while the low temperature regions expand.
If we see from the upper to the bottom panels,
the color is much bluer near the equatorial plane.
This reflects the fact that the higher black hole mass is,
the cooler becomes the accretion disk.
Because the flux is almost equal to the Eddington luminosity divided by surface area in the super-Eddington accretion disk, in other words, the disk temperature at a fixed $r/r_{\rm S}$ obeys $\sigma_{\rm SB} T^{4}_{\rm disk} \propto L_{\rm Edd}/(2\pi r^{2})\propto M^{-1}_{\rm BH}$
(see discussion in Kitaki et al. 2017).

The overall flow structure looks rather similar to those obtained by the previous study;
e.g., Kawashima et al. (2012) and Ogawa et al. (2017).
As was noted by Kitaki et al. (2017), we may distinguish three characteristic zones: 
accretion disk, funnel, and over-heated regions (see their Figure 2). 
The first one is the accretion disk located at and around the equatorial plane,
$r\sim r_{\rm in} - 1000r_{\rm S}$ and $\theta\sim30^{\circ}-90^{\circ}$
[see Model (a12) in Figure \ref{fig1}].
This disk is puffed up by the radiation pressure,
and gas falls toward the center by transporting the angular momentum outward via viscous process.

The second one is the funnel region located around the polar (rotational) axis, 
$r\sim r_{\rm in} - r_{\rm out}$ and $\theta\sim 0^{\circ} - 30^{\circ}$
[see Model (a12) in Figure \ref{fig2}].
The Thomson scattering optical depth of the funnel in the $z$-direction is $\tau_{\rm e}\sim 1$.
The funnel is characterized by high gas temperature, 
$k_{\rm B}T_{\rm funnel}\gtrsim 10{\rm keV}$ (see Figure \ref{fig2}),
and by a very fast velocity, $v_{r}\gtrsim 0.2c$.
The funnels in the first two columns from the left, which show models with low mass injection rates of
$\dot{M}_{\rm input}=300$ and $10^{3}$ ($\times L_{\rm Edd}/c^{2}$),
are widely extended from the polar direction to in the direction of $\theta\sim 45^{\circ}$, 
whereas the funnels in the third and fourth columns from the left, which show models with large mass injection rates of
$\dot{M}_{\rm input}= 5\times10^{3}$ and $10^{4}$ ($\times L_{\rm Edd}/c^{2}$) 
are rather narrow around the polar axis (see Figure \ref{fig2}).
This is because the funnel is collimated by the thickness of the puffed-up accretion disk
when the accretion rate is relatively high.

The third one is the over-heated region near the black hole at
$r\sim 5r_{\rm S}$ and $\theta\sim 45^{\circ}$
(see, e.g., Figure \ref{fig2} in Kitaki et al. 2017 for the details).
The gas temperature is very high ($\gtrsim 10{\rm keV}$) there.

\subsection{Mass accretion rate, outflow rate, and net flow rate}
\label{sec-flow-rate}
One of the most advantageous points in the present study is that we take a relatively large simulation 
box so that we could increase the angular momentum of the accretion material, as much as possible,
compared with those assigned in the previous studies.
This is a great advantage in investigating the super-Eddington accretion flow and outflow, since
we can achieve a quasi-steady state of inflow and outflow in a much wider spatial range than before.

We, here, discuss the properties of the simulated multi-dimensional accretion flow structure
in quasi-steady state in comparison with that calculated by the one-dimensional slim disk model.
For this purpose, it is essential to know to what extent a quasi-steady state is realized
and thus to calculate mass inflow rate and outflow rate at each radius.
In the present study, we calculate these rates according to
\begin{eqnarray}
  \dot{M}_{\rm in}(r)&=&\int_{4\pi}d\Omega ~r^{2}\rho(r,\theta){\rm min}\{v_{r}(r,\theta),0\},\\
  \dot{M}_{\rm out}(r)&=&\int_{4\pi}d\Omega ~r^{2}\rho(r,\theta){\rm max}\{v_{r}(r,\theta),0\}. \label{eq-outflow-rate}
\end{eqnarray}
Here, $\dot{M}_{\rm in} (<0)$ is the (time-averaged) mass inflow rate, and
$\dot{M}_{\rm out} (>0)$ is the (time-averaged) mass outflow rate.
(Note that we here calculate the mass outflow rate, irrespective of the outflow speed; that is,
we do not distinguish the outflow which can reach the infinity from the one that cannot.)
The net flow rate is then calculated by.
\begin{eqnarray}
  \dot{M}_{\rm net}(r)&=&\dot{M}_{\rm out}(r)+\dot{M}_{\rm in}(r).
\end{eqnarray}
By the quasi-steady flow we mean the flow, in which $\dot{M}_{\rm net}$ has 
no (or negligibly small) radial dependence.

\begin{figure}[h]
\begin{center}
  \includegraphics[width=80mm,bb=0 0 360 576]{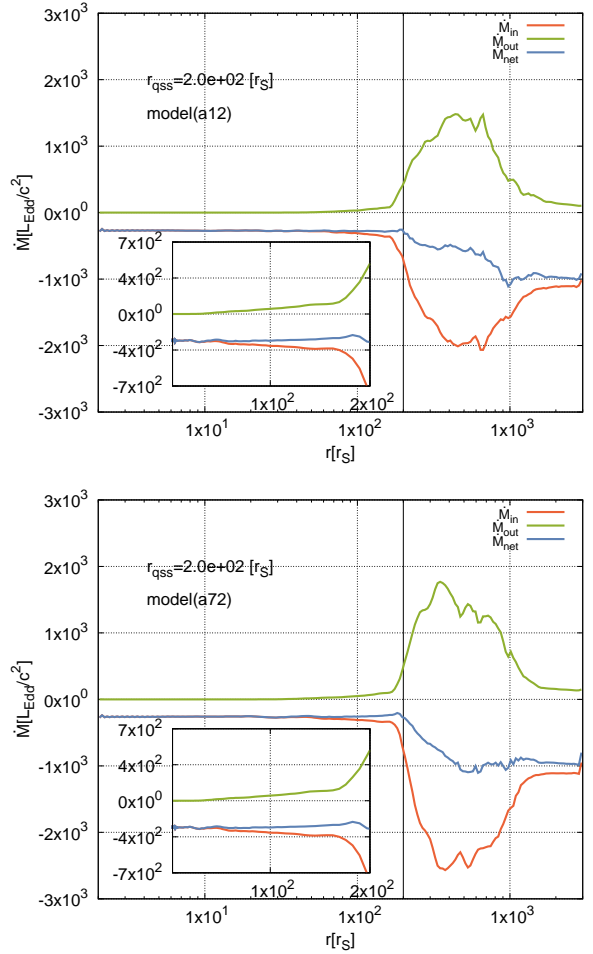}
\end{center}
\caption{The radial profiles of
  the (time-averaged) mass inflow rate, $\dot{M}_{\rm in}$ (red),
 of the (time-averaged) mass outflow rate, $\dot{M}_{\rm out}$ (green), and
 of the net flow rate, $\dot{M}_{\rm net}\equiv \dot{M}_{\rm in} + \dot{M}_{\rm out}$ (blue) 
  for Model (a12) and Model (a72) in the upper and lower panels, respectively. 
  Note that the horizontal scale is in the logarithm.   
  The vertical black lines at $r\sim 200r_{\rm S}$ indicate the radius of $r_{\rm qss}$, 
  inside which the flow is in a quasi-steady state (see text for the precise definition).
  The inset in each panel shows the same but in a narrower spatial range of $r\leq 200r_{\rm S}$,
  in which the horizontal axis is now in the linear scale.
}
\label{fig-mdot-small}
\end{figure}

Figure \ref{fig-mdot-small} shows typical examples of the radial distributions
of the inflow rate, the outflow rate, and the net rate for Models (a12) and (a72).
(We wish to note that other models show similar results.)
We see that $\dot{M}_{\rm net}$ is nearly constant in a spatial range of $r\sim 2-200r_{\rm S}$.
We also notice that both of 
the absolute value of the mass inflow rate 
and the mass outflow rate 
increase significantly beyond the radius of $r\sim200 r_{\rm S}$.
Such features can be understood, since 
the injected gas from the outer boundary at $r=r_{\rm out}$ is once accumulated around 
the Keplerian radius, $r\sim r_{\rm K}$, due to the angular momentum barrier.
We thus conclude that a quasi-steady flow is achieved in the range of $r\sim 2-200r_{\rm S}$.

Let us define the quasi-steady radius, $r_{\rm qss}$,
as the radius, inside which a quasi-steady state realizes; i.e.,
the net flow rate is constant, $\dot{M}_{\rm net}(r)\propto r^{0}$.
Technically, we evaluate $r_{\rm qss}$ in the following way.
\begin{enumerate}
\item We first give a guess value of $r_{\rm qss}$,
say, $r_{\rm qss}=100 r_{\rm S}$.
\item We search for the radial mesh index ${i_{\rm qss}}$ such that
the following inequality holds; $r_{i_{\rm qss}}\leq r_{\rm qss} < r_{i_{\rm qss}+1}$.
\item The mean net flow rate, $\average{\dot{M}_{\rm net}}$, and its standard deviation, 
$\sigma_{\rm net}$, are calculated by averaging the net flow rate over the range between
$r=r_{\rm in}$ and $r_{i_{\rm qss}+1}$; that is.
  \begin{eqnarray}
   \average{\dot{M}_{\rm net}}&\equiv&
      \frac{1}{i_{\rm qss}}\sum_{i=1}^{i_{\rm qss}}\dot{M}_{\rm net}(r_{i}),\\
   \sigma_{\rm net}&\equiv&
      \sqrt{\frac{1}{i_{\rm qss}}\sum_{i=1}^{i_{\rm qss}}
           \left[\dot{M}_{\rm net}(r_{i})-\average{\dot{M}_{\rm net}}.\right]^{2}}.
  \end{eqnarray}
\item If the relationship $\left|\dot{M}_{\rm net}(r_{i_{\rm qss}+5})\right|
   \geq\left|\average{\dot{M}_{\rm net}}\right|+1.5\sigma_{\rm net}$
  holds for the first time, we define the radius $r_{\rm qss}$ as
  \begin{eqnarray}
    r_{\rm qss}&\equiv& (r_{i_{\rm qss}}+r_{i_{\rm qss}+1})/2,
  \end{eqnarray}
 and end the loop.
  Otherwise, we repeat the same procedure from the first step (1.) but by adding 1 to $i_{\rm qss}$.
\end{enumerate}

The quasi-steady radius, $r_{\rm qss}$, as is indicated in Figure \ref{fig-mdot-small},
and the black hole accretion rate, 
$\dot{M}_{\rm BH}\equiv |\dot{M}_{\rm in}(r=r_{\rm in})|$, are listed in table \ref{tab-mdot}.
We understand from table \ref{tab-mdot}
that a quasi-steady state realizes inside $(1-2)\times 10^2 r_{\rm S}$) and
that $\dot{M}_{\rm BH}$ exceeds several tens of $L_{\rm Edd}/c^2$, meaning that
the super-Eddington accretion flow is actually occurring (see, e.g. Watarai et al. 2001).

\begin{table}[h]
  \tbl{Net mass accretion rates in quasi-steady regions.}{
    \begin{tabular}{ccc}
      \hline     
      model & $\dot{M}_{\rm BH}$[$L_{\rm Edd}/c^{2}$] & $r_{\rm qss}[r_{\rm S}]$\\
      \hline
      (a11) & $1.4\sisuu{2}$ & $ 1.2\sisuu{2}$ \\
      (a12) & $2.8\sisuu{2}$ & $ 2.0\sisuu{2}$ \\
      (a13) & $7.9\sisuu{2}$ & $ 1.6\sisuu{2}$ \\
      (a14) & $8.3\sisuu{2}$ & $ 1.6\sisuu{2}$ \\
      (a15) & $5.5\sisuu{3}$ & $ 1.6\sisuu{2}$ \\\hline
      (a41) & $1.4\sisuu{2}$ & $ 9.9\sisuu{1}$ \\      
      (a42) & $2.7\sisuu{2}$ & $ 1.9\sisuu{2}$ \\
      (a43) & $7.5\sisuu{2}$ & $ 1.6\sisuu{2}$ \\
      (a44) & $9.4\sisuu{2}$ & $ 1.5\sisuu{2}$ \\
      (a45) & $5.5\sisuu{3}$ & $ 1.6\sisuu{2}$ \\\hline
      (a71) & $1.4\sisuu{2}$ & $ 9.6\sisuu{2}$ \\      
      (a72) & $2.7\sisuu{2}$ & $ 2.0\sisuu{2}$ \\
      (a73) & $8.0\sisuu{2}$ & $ 1.6\sisuu{2}$ \\
      (a74) & $9.3\sisuu{2}$ & $ 1.5\sisuu{2}$ \\
      (a75) & $5.6\sisuu{3}$ & $ 1.6\sisuu{2}$ \\\hline
  \end{tabular}}
  \begin{tabnote}
    The time-averaged mass accretion rate onto the black hole, ${\dot{M}_{\rm BH}}$,
   and quasi-steady radius, $r_{\rm qss}$, inside which the quasi-steady state realizes
    (see Figure \ref{fig-mdot-small}).
    Note that $\dot{M}_{\rm BH}\equiv |\dot{M}_{\rm in}(r=r_{\rm in})|$.
  \end{tabnote}
  \label{tab-mdot}     
\end{table}

\subsection{Scaling relations of flow structure}
\begin{figure*}[h]
\begin{center}
  \includegraphics[width=160mm,bb=0 0 720 432]{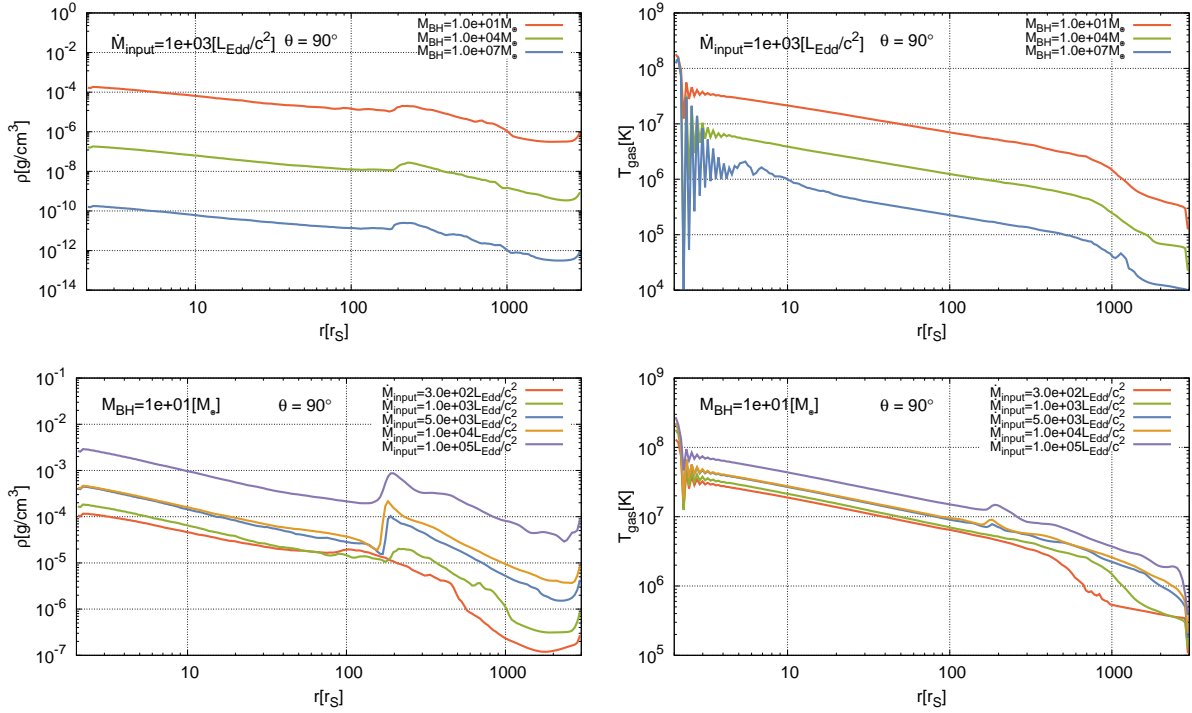}
\end{center}
\caption{
  [Top] The radial distributions of the mass density $\rho$ (left panels) and temperature 
   $T_{\rm gas}$ (right panels) on the equatorial plane for various black hole masses of
   $M_{\rm BH}=10$ (red), $10^{4}$ (green), and $10^{7}M_{\odot}$ (blue), respectively, 
   but for a fixed $\dot{M}_{\rm input}=10^{3}L_{\rm Edd}/c^{2}$.
  [Bottom] Same as the top panels for a variety of mass injection rates of
  $\dot{M}_{\rm input}=300$ (red), $10^{3}$ (green), $5\times10^{3}$
    (blue), $10^{4}$ (yellow), and $10^{5}L_{\rm Edd}/c^{2}$(purple) for a fixed $M_{\rm BH}=10M_{\odot}$.
}
\label{fig-mix1}
\end{figure*}

Rather systematic variations found in the panels of Figures \ref{fig1} and \ref{fig2} 
indicate the existence of simple scaling laws for the functional dependences
of the density and temperature distributions on $M_{\rm BH}$ and $\dot{M}_{\rm BH}$,
To demonstrate that it is really the case, we plot in
Figure \ref{fig-mix1} the radial distributions of mass density $\rho$ and gas temperature $T_{\rm gas}$
on the equatorial plane for several models.

First of all, we notice that each line is nearly straight inside the quasi-steady radius, 
meaning that density and temperature are power-law functions of radius there. 
Next, we find a roughly constant interval between each line. 
We thus expect the following universal scaling relations to hold for any physical quantities, $f$;
\begin{eqnarray}
  f&=&A_{f}\times \left(\frac{M_{\rm BH}}{M_{\odot}}\right)^{a}\left(\frac{\dot{M}_{\rm BH}}{L_{\rm Edd}/c^{2}}\right)^{b}\left(\frac{r}{r_{\rm S}}\right)^{c},
\end{eqnarray}
where $A_{f}$, $a$, $b$, and $c$ are numerical constants that depend on the physical quantities $f$
but are independent of $M_{\rm BH}, \dot{M}_{\rm BH}$ and $r$. The best fit values on the equatorial plane are:
\begin{eqnarray}
  \rho&=& (9.08\pm 1.25) \sisuu{-6} [{\rm g~cm^{-3}}] \nonumber \\ 
  &&\times\left(\frac{M_{\rm BH}}{M_{\odot}}\right)^{-1.00}
    \left(\frac{\dot{M}_{\rm BH}}{L_{\rm Edd}/c^{2}}\right)^{1.04}
    \left(\frac{r}{r_{\rm S}}\right)^{-0.73},  \label{eq-dep-rho}\\
  T_{\rm gas}&=& (3.85\pm 0.33)\sisuu{7} [{\rm K}] \nonumber \\ 
   &&\times\left(\frac{M_{\rm BH}}{M_{\odot}}\right)^{-0.24}
     \left(\frac{\dot{M}_{\rm BH}}{L_{\rm Edd}/c^{2}}\right)^{0.24}
     \left(\frac{r}{r_{\rm S}}\right)^{-0.54},  \label{eq-dep-tgas}\\
  E_{0}&=&(2.36 \pm 0.14) \sisuu{15} [{\rm erg~cm^{-3}}] \nonumber \\ 
   &&\times\left(\frac{M_{\rm BH}}{M_{\odot}}\right)^{-1.00}
     \left(\frac{\dot{M}_{\rm BH}}{L_{\rm Edd}/c^{2}}\right)^{1.02}
     \left(\frac{r}{r_{\rm S}}\right)^{-1.73},  \label{eq-dep-rade}\\
  v_{r}&=& (-0.36\pm 0.01) [c]\nonumber\\  
   &&\times\left(\frac{M_{\rm BH}}{M_{\odot}}\right)^{0.00}
     \left(\frac{\dot{M}_{\rm BH}}{L_{\rm Edd}/c^{2}}\right)^{0.02}
     \left(\frac{r}{r_{\rm S}}\right)^{-1.11},  \label{eq-dep-vr}\\
  v_{\phi}&=& (0.81\pm 0.02)[c]\nonumber\\ 
  &&\times\left(\frac{M_{\rm BH}}{M_{\odot}}\right)^{0.00}
    \left(\frac{\dot{M}_{\rm BH}}{L_{\rm Edd}/c^{2}}\right)^{0.01}
    \left(\frac{r}{r_{\rm S}}\right)^{-0.60}.  \label{eq-dep-vphi}
\end{eqnarray} 
We also calculate the standard deviations of $a$, $b$, and $c$, confirming
that they are sufficiently small, much less than unity.
In the next section we will compare these scaling laws with those by the slim disk model.

\subsection{The effective temperature}
In the last subsection we investigate the functional dependence of a more directly observable 
quantity; i.e., the effective temperature, $T_{\rm eff}$.
For obtaining the effective temperature distributions,
we solve
the grey radiative transfer equation with isotropic scattering;
\begin{eqnarray}
  \mu\frac{dI}{dz}&=&\frac{1}{4\pi}\epsilon_{\rm ff} - (\alpha_{\rm ff} + \rho\kappa_{\rm es}) I + \rho\kappa_{\rm es}J. \label{eq-rad-trans}
\end{eqnarray}
Here, $I$ is the specific intensity,
$\epsilon_{\rm ff}=1.4\times10^{-27}T^{1/2}(\rho/m_{\rm p})^{2}[{\rm erg~sec^{-1}~cm^{-3}}]$ ($m_{\rm p}$ is the proton mass) is the emissivity,
$\alpha_{\rm ff}=1.7\times10^{-25}T^{-7/2}(\rho/m_{\rm p})^{2}[{\rm cm^{-1}}]$
is the absorption coefficient,
$\kappa_{\rm es}=\sigma_{\rm T}/m_{\rm p}$ and $\sigma_{\rm T}$ is the Thomson cross section,
and $J=(1/4\pi)\int I d\Omega = cE_{0}/(4\pi)$ is the mean intensity,
$\mu$ is the direction cosine,
respectively.
In the present study we fix $\mu\equiv 1$ for simplicity.

We solve equation (\ref{eq-rad-trans}) numerically in the $z$-direction
at a fixed cylindrical radius, $R=r\sin\theta$.
Using the value $E_{0}(\propto J)$ calculated from 2D-RHD code,
the solution is,
\begin{eqnarray}
  I(\tau)&=&\int_{\tau}^{\tau(-z_{\rm max})}S(\tau^{\prime})e^{-(\tau^{\prime}-\tau)}d\tau^{\prime},\label{eq-intensity}
\end{eqnarray}
where
\begin{eqnarray}
  S(\tau)&\equiv&\frac{1}{4\pi} \frac{\epsilon_{\rm ff}+\rho\kappa_{\rm es}cE_{0}}{\alpha_{\rm ff}+\rho\kappa_{\rm es}},
\end{eqnarray}
and
\begin{eqnarray}
  \tau(z)&\equiv&\int^{z_{\rm max}}_{z}(\alpha_{\rm ff}+\rho\kappa_{\rm es})dz.
\end{eqnarray}
We set $z_{\rm max}$ to be the outer boundary of the simulation box; that is,
\begin{equation}
 z_{\rm max}\equiv\sqrt{r_{\rm out}^{2}-R^{2}}. \label{zmax}
\end{equation}

\begin{figure}[t!]
\begin{center}
  \includegraphics[width=80mm,bb=0 0 360 432]{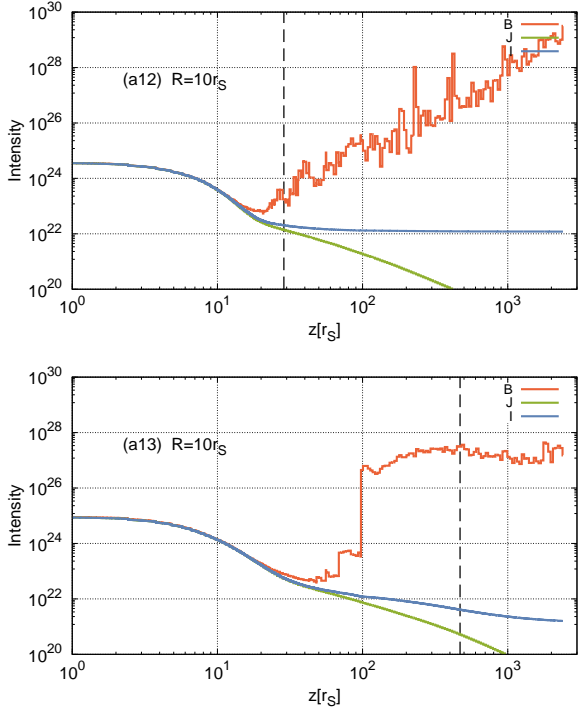}
\end{center}
\caption{The specific intensity $I$ (blue, equation \ref{eq-intensity}),
  the blackbody $B=\sigma_{\rm SB}T_{\rm gas}^{4}/\pi$ (red),
  and the mean intensity $J=cE_{0}/4\pi$ (green) at radius $R=10r_{\rm S}$,
  for Models (a12) and (a13) in the top and bottom panels, respectively.
  The vertical black dashed line means the position of $\tau(z)=1$.
  The optical depth at $z=100r_{\rm S}$ in the bottom panel is $\tau(z=100r_{\rm S})\sim 2$.
}
\label{fig-intensity}
\end{figure}

Figure \ref{fig-intensity} shows the solution (\ref{eq-intensity})
at radius $R=10r_{\rm S}$ for Models (a12) and (a13).
Let us examine the case of Model (a12) first (see the top panel).
The specific intensity $I$ near the equatorial plane ($z\sim 0$) is equal to
the blackbody $B=\sigma_{\rm SB}T_{\rm gas}^{4}/\pi$
and the mean intensity $J=cE_{0}/4\pi$.
This is because the optical depth is large within the accretion disk.
The larger $z$ is,
the lower gas temperature $T_{\rm gas}$ becomes, and so does the specific intensity
$I=B\propto T^{4}_{\rm gas}$ 
at around $z\sim 1r_{\rm S} - 10r_{\rm S}$.

In the middle region of $z\sim 10r_{\rm S} - 30r_{\rm S}$,
the layer is marginally optically thick ($\tau\gtrsim 1$)
and so the specific intensity $I$ does no longer match the blackbody intensity $B$.
This is because a decrease of intensity by scattering of photons out of the ray ($-\rho\kappa_{\rm es}I$) 
is dominant over an increase of intensity by scattering of photons into the ray ($\rho\kappa_{\rm es}J$).
We confirm that other terms ($\epsilon_{\rm ff}/4\pi$ and $-\alpha_{\rm ff} I$) are of minor importance in equation (\ref{eq-rad-trans}).
Hence,
the intensity $I$ 
should become weaker and weaker with an increase of $z$ until $z\sim 30r_{\rm S}$,
where $\tau\sim1$ holds. We can say that this layer corresponds to a photosphere. 

Above the photosphere the intensity $I$ stays roughly constant, since
radiation hardly interacts with gas there.
We can thus approximate the specific intensity reaching a distant observer to be that 
at the outer simulation boundary, $z=z_{\rm max}$ [see equation (\ref{zmax})].
We can then calculate the effective temperature $T_{\rm eff}$ at radius $R$ by
\begin{equation}
  T_{\rm eff}(R)\sim [\pi I(z=z_{\rm max})/\sigma_{\rm SB}]^{1/4}. 
\label{eq-temp-eff}
\end{equation}

Let us next examine the case of Model (a13) (see the bottom panel in Figure \ref{fig-intensity}).
The specific intensity $I$ behaves in a similar way to that of Model (a12),
but the $z$-dependence of intensity is not exactly the same between them.
The specific intensity $I$ is roughly constant above the photosphere in Model (a12), whereas 
it still decreases slowly even above the photosphere at $z \sim 300 r_{\rm S}$ in Model (a13).
(Note that the higher the mass injection rate is, the larger becomes the scale-height.)
We expect that $I$ will stay nearly constant above $\sim 3000 r_{\rm S}$, 
although this is not numerically confirmed.
In this paper, therefore, we calculate $T_{\rm eff}(R)$ by inserting the intensity
at $z=z_{\rm max}$ into equation (\ref{eq-temp-eff}).
We should note that this $T_{\rm eff}(R)$ is likely to be overestimated.
We confirm that the shape of the intensity curve (i.e. $I, J, B$ in Figure \ref{fig-intensity})
looks the same at different $R$ in the model(a12).
We also confirm that the intensity curve in all models
presents the same behaviour as
in Figure \ref{fig-intensity}.

\begin{figure}[ht]
\begin{center}
  \includegraphics[width=80mm,bb=0 0 360 432]{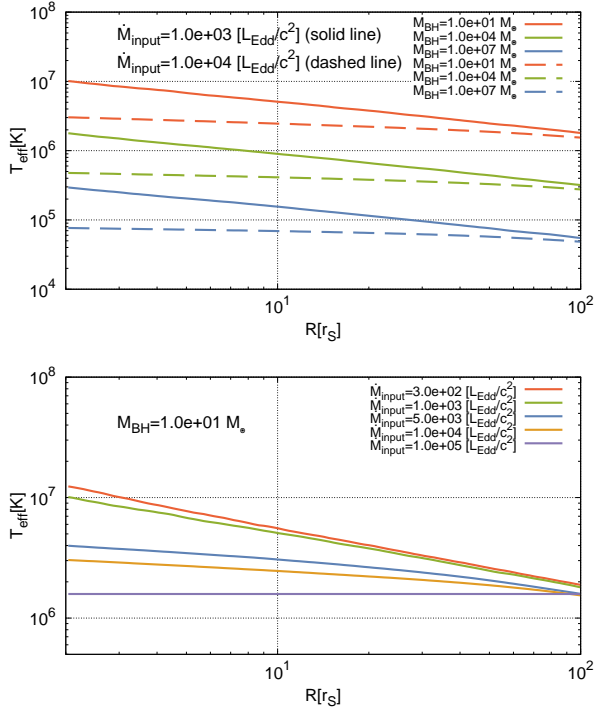}
\end{center}
\caption{
  The radial profiles of the effective temperature $T_{\rm eff}(R)$
  for various black hole masses, $M_{\rm BH}$ (top panel),
  and for various mass injection rates, $\dot{M}_{\rm input}$ (bottom panel).
  Note that the effective temperature distribution for high mass injection rates of,
  $\dot{M}_{\rm input}=5\times 10^{3}, 10^{4}$ and $10^{5} (L_{\rm Edd}/c^{2})$
  may not be so accurate due to a limited size of the computational box (see text).
}
\label{fig-temp-eff}
\end{figure}

We calculate the effective temperature as a function of $R$, $T_{\rm eff}(R)$,
for various models and plot them in Figure \ref{fig-temp-eff}.
As one can see in the top panel of Figure \ref{fig-temp-eff},
the effective temperature is proportional to $T_{\rm eff}\propto M_{\rm BH}^{-1/4}$,
as long as $\dot{M}$ is moderately large, $\dot{M} \lesssim 10^3 L_{\rm Edd}/c^2$.
We confirm that this relation holds for other models with different $M_{\rm BH}$.
As $\dot {M}$ increases, however, the effective temperature ($T_{\rm eff}$) profile 
obviously becomes flatter and the innermost temperature significantly drops
(see the bottom panel in Figure \ref{fig-temp-eff}).

We wish to note again that the effective temperature calculated for models with $\dot{M}_{\rm input}=5\times10^{3}, 10^{4}$ and $10^{5}L_{\rm Edd}/c^{2}$ is likely to be overestimated. This is because the photosphere is not in the numerical box size at larger distances for the high mass accretion rate model. We think that the flatter profile at high accretion rates is linked to the overestimate of $T_{\rm eff}$ due to the numerical box size, an effect which increases at larger distances from the black hole.

Why does the effective temperature decrease as the mass accretion rate increases?
The multi-dimensional photon-trapping effect may be a reason (Ohsuga et al. 2002).
To demonstrate that this is the case, we calculate three types of radiation advection,
inward, outward, and net advection rates of radiation energy, as functions of $r$ by
\begin{eqnarray}
L_{\rm in}(r)&=&\int_{4\pi}d\Omega~{\rm min}\{0, r^{2}\left[F_{r}+E_{0}v_{r}\right] \}~~~~~~ <0,\label{rad-adv1}\\
L_{\rm out}(r)&=&\int_{4\pi}d\Omega~{\rm max}\{0,  r^{2}\left[F_{r}+E_{0}v_{r}\right] \}~~~~~~ >0,\label{rad-adv2}
\end{eqnarray}
and
\begin{eqnarray}
L_{\rm net}&=&L_{\rm in}+L_{\rm out}\label{rad-adv3}.
\end{eqnarray}

\begin{figure}[ht]
\begin{center}
  \includegraphics[width=80mm,bb=0 0 360 432]{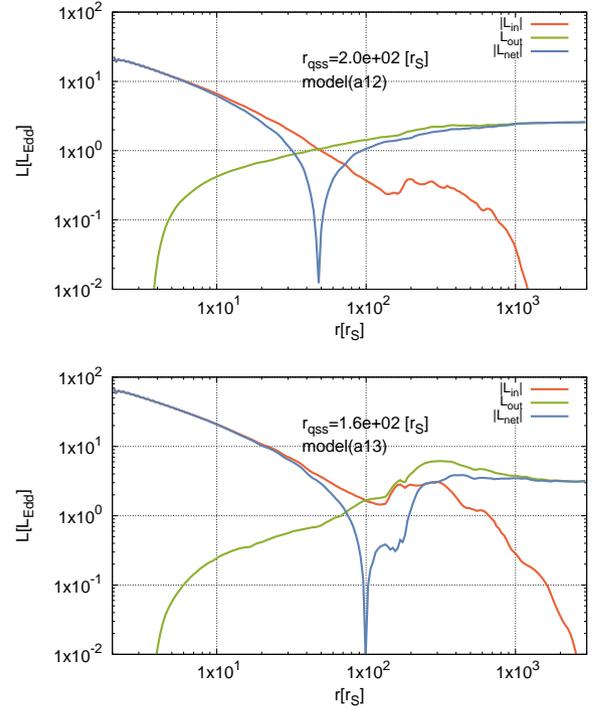}
\end{center}
\caption{
  The radial profile of the inward and outward advection rates of radiation 
  [see equations (\ref{rad-adv1})-(\ref{rad-adv3})].
  In both panels the inward advection of radiation (i.e., photon trapping) dominates
  inside the radius of several tens of $r_{\rm S}$.
}
\label{fig-rad-adv}
\end{figure}

Figure \ref{fig-rad-adv} shows $\left|L_{\rm in}\right|$, $\left|L_{\rm out}\right|$, and $\left|L_{\rm net}\right|$
as functions of radius.
This figure clearly shows that
the inward advection of radiation energy is dominant over the outward advection near the black hole, 
and that the higher the mass accretion rate is, the larger becomes $|L_{\rm in}|$. 
We confirm that $|L_{\rm in}|$ is about 3.1 times larger in Model (a13) than in Model (a12) 
at $r=10r_{\rm S}$.
Thus, the radiative flux emerging from the innermost part is significantly reduced,
as $\dot {M}$ increases.
This is just a qualitative argument and its quantitative assessment is left as a future work.
Discussion regarding to what extent
the boundary conditions, the spatial resolution, and the computational size affect
the surface temperature is also future issues.

Finally, we here give scaling laws of the effective temperature only for the cases with
$\dot{M}_{\rm input}=300$ and $10^{3} (L_{\rm Edd}/c^{2})$, that is,
\begin{equation}
T_{\rm eff}=(2.93\pm0.01)\sisuu{7}[{\rm K}]\nonumber\\
    \times\left(\frac{M_{\rm BH}}{M_{\odot}}\right)^{-0.25}\left(\frac{r}{r_{\rm S}}\right)^{-0.47}, \label{eq-temp-eff-fit1}\\
\end{equation}
for $\dot{M}_{\rm input} =300 L_{\rm Edd}/c^{2}$ and
\begin{equation}
T_{\rm eff}=(2.47\pm0.01)\sisuu{7}[{\rm K}]\nonumber\\
   \times\left(\frac{M_{\rm BH}}{M_{\odot}}\right)^{-0.25}\left(\frac{r}{r_{\rm S}}\right)^{-0.44}, \label{eq-temp-eff-fit2}
\end{equation}
  for $\dot{M}_{\rm input} =10^3 L_{\rm Edd}/c^{2}$, respectively.
As we will see in the next section, these functional dependences on $M_{\rm BH}$ and $r$
are in good agreement with those of the slim disk model.

The reason why we do not show the scaling laws for models with higher mass injection rates 
is that the location of the photosphere is very close to the outer boundary of the calculation box
so that the effective temperature calculations may not be so reliable.
We need an even larger computational box size in a future study.

\section{Discussion}
\subsection{Comparison with the slim disk model} 
We obtained the fitting formulas of super-Eddington accretion disk for the first time
by the systematic study of the scaling relations produced by numerical simulations.
It will be interesting to examine how well one-dimensional slim disk model can reproduce
our simulation results. This is a very fundamental issue but surprisingly it has been poorly investigated 
in past simulation studies due probably to the limited spatial resolution.
We are not at a position to answer to this question.

For comparison purpose, we use the data on the equatorial plane by Watarai (2006);
\begin{eqnarray}
  \rho^{\rm slim}(r)&=& 1.2\sisuu{-5}[{\rm g~cm^{-3}}]\nonumber\\ 
  &&\times\left(\frac{M_{\rm BH}}{M_{\odot}}\right)^{-1}
  \left(\frac{\dot{M}}{L_{\rm Edd}/c^{2}}\right)
  \left(\frac{r}{r_{\rm S}}\right)^{-3/2},\label{eq-slim-rho}\\
  T_{\rm gas}^{\rm slim}(r)&=& 5.3\sisuu{7}[{\rm K}]\nonumber\\
  &&\times\left(\frac{M_{\rm BH}}{M_{\odot}}\right)^{-1/4}
  \left(\frac{\dot{M}}{L_{\rm Edd}/c^{2}}\right)^{1/4}
  \left(\frac{r}{r_{\rm S}}\right)^{-5/8},\label{eq-slim-tgas}\\
  T_{\rm eff}^{\rm slim}(r)&=& 4.5\sisuu{7}[{\rm K}]
  \left(\frac{M_{\rm BH}}{M_{\odot}}\right)^{-1/4} 
  \left(\frac{r}{r_{\rm S}}\right)^{-1/2},\label{eq-slim-teff}\\
  v_{r}^{\rm slim}(r)&=& -1.1\sisuu{-1}[c]
  \left(\frac{r}{r_{\rm S}}\right)^{-1/2},\label{eq-slim-vr}\\
  v_{\phi}^{\rm slim}(r)&=& 7.1\sisuu{-1}[c]
  \left(\frac{r}{r_{\rm S}}\right)^{-1/2}.\label{eq-slim-vphi}
\end{eqnarray}
Here, $\alpha=0.1$ and $\gamma=4/3$ are set in these equations.

Let us compare these functional dependences with those obtained by our RHD simulations
[see equations (\ref{eq-dep-rho})-(\ref{eq-dep-vphi})].
We soon notice that the dependences on the black hole mass, the mass accretion rate,
and radius of each physical quantity are in reasonable agreement between the two.
Two important exceptions are
the radial dependences of the mass density $\rho$ and of the radial velocity $v_{r}$.
In the former, in particular, we find $\rho \propto r^{-1.5}$ according to the slim disk model
while the density profile is much shallower; $\rho \propto r^{-0.5}$.
Why is so different?

It will be interesting to note in this respect that similar discrepancy had been found
in the simulation study of RIAF (radiatively inefficient accretion flow). 
Igumenshchev et al. (1999) were the first to demonstrate by their hydrodynamic simulations
that pure ADAF (advection-dominated accretion flow) appears when the $\alpha$ viscosity
parameter is relatively large ($\alpha \sim 0.1$), leading to a steep density profile, $\rho \propto r^{-1.5}$,
while convection arises when the $\alpha$ is small ($\alpha \sim 0.01$),
giving rise to a much flatter density profile, $\rho \propto r^{-0.5}$.
The latter type of flow is sometimes called as convection-dominated accretion flows (CDAF).
Machida et al. (2001) examined the radial density profile by performing 3-D MHD simulations
and confirmed the existence of large-scale circulation.
The density profile is accordingly flatter; $\rho\propto r^{-0.5}$, $v_{r}\propto r^{-1.3}$.
These previous study considered in the case of the low mass accretion rate ($\dot{M}_{\rm BH}\leq L_{\rm Edd}/c^{2}$).
Here, we stress that even if we set $\alpha=0.1$, the convection occurs in the present study
with high mass accretion rate ($\dot{M}_{\rm BH}\geq L_{\rm Edd}/c^{2}$),
and the radial profile of the mass density is $\rho\propto r^{-0.73}$.

We thus checked the simulation movies of the RHD simulations, and confirmed
the occurrence of large-scale circulation (or convection) within the accretion disk
(see also Ohsuga et al. 2005).
The two-dimensional velocity map in the $R$-$z$ plane also supports the CDAF type flow.
We may thus tentatively conclude that the flatter density profile in our RHD simulation data
could be the results of the convection, which is not properly considered in the slim disk model.
We should then note that the density profile may depend on the adopted $\alpha$ value.
This point needs to be checked in future radiation-MHD (R-MHD) simulations.
Note that the $v_r$ profile is determined by the quasi-steady condition;
$\dot{M} = -2\pi r v_{r} \rho H  \sim$ const.
with $H (\sim r)$ being the scale-height of the inflow disk.

The effective temperature profile in equations (\ref{eq-temp-eff-fit1}) and (\ref{eq-temp-eff-fit2}) also agrees well with that of the slim disk model [see equation (\ref{eq-slim-teff})].
The effective temperature of the standard disk (Shakura \& Sunyaev 1973) is proportional to $T_{\rm eff}^{\rm standard}\propto r^{-3/4}$.
But, when the mass accretion rate becomes higher ($\dot{M}\gtrsim L_{\rm Edd}/c^{2}$), 
the radial dependence of the effective temperature becomes flatter than that of the standard disk 
[see equation (\ref{eq-slim-teff})].
This is because the photon-trapping effects.
We can also understand the behavior of the effective temperature
by this relationship $T_{\rm eff}^{4}\propto F\propto L_{\rm Edd}/(2\pi r^{2})$
(Kato et al. 2008).

In our RHD simulations,
$T_{\rm gas} \sim T_{\rm rad}\propto E_{0}^{1/4}$ is established from equation (\ref{eq-dep-tgas}), (\ref{eq-dep-rade}),
which means that the accretion disk is optically thick.
This relation is consistent with one of the assumptions needed for constructing the slim disk model.

\subsection{Comparison with previous simulations}
In this subsection we compare our RHD simulation results with those 
of previous simulations to stress what is new in the present study.

Hashizume et al. (2015) performed the RHD simulations using the same code used in Ohsuga et al. (2005),
but the computational box was set to be larger ($r_{\rm out}=5000r_{\rm S}$).
The mass injection rate in Hashizume et al. (2015) was $\dot{M}_{\rm input}=10^{3}L_{\rm Edd}/c^{2}$,
and the initial Keplerian radius of the injected gas was $r_{\rm K}=100r_{\rm S}$.
The important difference between our study and Hashizume et al. (2015)
lies in that Compton effects were not taken into account in their simulations.
According to their Figure 4, 
the net flow rate is roughly constant in radius at $r\lesssim 100r_{\rm S}$.
The outflow rate is negligible near the black hole (at $r\lesssim60r_{\rm S}$), 
while it is substantial in the outside region at $r\gtrsim 60r_{\rm S}$.
Such a separation of the innermost region without outflow 
and the outer region with significant outflow is also observed in our simulation data
(see Figure \ref{fig-mdot-small}), although the separating radius 
(i.e., quasi-steady radius, $r_{\rm qss}$) is much less in their simulations.
This is because
the outflowing gas density becomes significantly lower
when we include the effects of Compton cooling as shown in Figure 1 (c) and (d)
in Kawashima et al. (2009).


S\c{a}dowski et al. (2015) performed GR-R-MHD simulation of super-Eddington accretion flow
onto a $10M_\odot$ black hole for various simulation parameters (black hole spin, 
initial magnetic field strength and configurations, etc).
According to their Figure 6, 
a quasi-steady state is achieved inside $30r_{\rm S}$,
while outflow mainly emerges outside $\sim 10r_{\rm S}$.
Although the trend that the outflow hardly emerges from the black hole vicinity is consistent with 
our simulation results, the radial extent, in which outflow is negligible, is 
significantly narrower in their simulations.
This difference seems to arise in the fact that they adopted a small radius 
for the centroid location of the initial torus ( $\sim 21r_{\rm S}$).

Let us next compare our results and theirs in terms of the velocity profiles.
The azimuthal velocity $v_{\phi}$ is grossly sub-Keplerian (see their Figure 13),
which is consistent with our result.
The mass density weighted and azimuthally averaged radial velocity ($\average{v_{r}}_{\theta}$)
approximately obeys the relationship of $\average{v_{r}}_{\theta}\propto r^{-2}$ 
at $r\lesssim 30r_{\rm S}$ (see their Figure 16), which is much stepper than our results;
$\average{v_{r}}_{\theta}\propto r^{-1.25}$.
This radial dependence is very close to that on the equatorial plane;
 $v_{r}\propto r^{-1.11}$ [see equation (\ref{eq-dep-vr})].
(This similar radial dependence is not so surprising, since mass density is at maximum 
at around the equatorial plane.)
To conclude,
the radial dependence of the accretion velocity on the equatorial plane in S\c{a}dowski et al. (2015)
is very different from our results ($v_{r}\propto r^{-1.11}$).
This discrepancy could arise due to different treatments of disk viscosity
(or magnetic processes).
We adopted the $\alpha$-viscosity model, whereas they solved the MHD processes
in the axisymmetric geometry with a sub-grid magnetic dynamo.
Again, full three-dimensional radiation-MHD (R-MHD) simulations are necessary to settle this issue.

\subsection{Why is outflow so weak from the innermost region?}

\begin{figure}[h]
\begin{center}
  \includegraphics[width=80mm,bb=0 0 288 288]{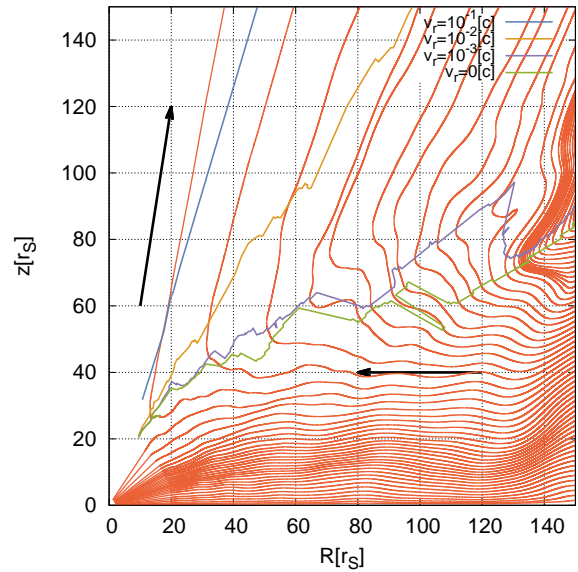}
\end{center}
\caption{The streamlines (red lines)
  and the loci of constant radial velocities of
  $v_{r}=10^{-1}$ (blue line), $10^{-2}$ (yellow line),
  $10^{-3}$ (purple line), and $0[c]$ (green line)
  for Model (a12).
  The black arrows indicate the direction of gas flow along the streamlines.
}
\label{fig-streamline1}
\end{figure}

\begin{figure}[h]
\begin{center}
  \includegraphics[width=80mm,bb=0 0 360 216]{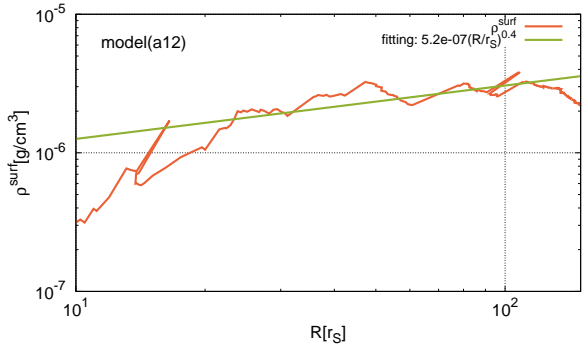}
\end{center}
\caption{
  Radial distribution of mass density (red) at the surface of the inflow
  region and its best-fit line (green) at $R=10-100[r_{\rm S}]$
  for Model (a12).
  Here, by the {\lq}surface{\rq} we mean the location of vanishing radial 
  velocity, $v_{r}=0$; namely, $v_r < 0$ (or $v_r > 0$) below (above) the surface.
  (see the green line in Figure \ref{fig-streamline1}).
}
\label{fig-streamline2}
\end{figure}

 From Figure \ref{fig-mdot-small}, we understand that
the inflow rate is roughly constant in radius, while outflow rate is very small near the black hole. 
This feature is in good agreement with the slim-disk formulation
(since mass-flow rate is assumed to be constant)
but does not quantitatively agree with the previous RHD simulation results.
How can we understand this?

The small outflow rate could be due to (1) low density at the launching point of outflow 
(i.e., inflow surface), (2) slow outflow velocity, or (3) combination thereof
(see equation \ref{eq-outflow-rate}).
To examine which is the case, we plot the gas streamlines, as well as
the radial velocity contour lines $v_{r}=0, 10^{-1}, 10^{-2},$ and $10^{-3}[c]$, 
for Model (a12) in Figure \ref{fig-streamline1}.
We readily understand that
outflow emerges even from near the black hole ($r\lesssim 100r_{\rm S}$)
and that the outflow speed is not small; $|v_r| \gtrsim 0.1 c$. 
When we follow each streamline near the black hole, we see that
the outflow is accelerated up to nearly the speed of light.
Thus, we conclude that
the mass density at the flow surface should be very small near the black hole
to account for the small outflow rates.

The mass density $\rho^{\rm surf}$ at the flow surface for Model (a12) is plotted 
as a function of radius in Figure \ref{fig-streamline2}.
Here, by the surface we mean the places where the radial velocity vanishes;
$v_{r}=0$. (There are the places where outflow is launched.)
The best-fit line (in the log-log plot) in the range of $R=10-100r_{\rm S}$ is 
$\rho^{\rm surf}\propto R^{0.4}$. That is, density is decreasing as matter accretes.
This supports that the gas density at the outflow launching site is indeed very small.

To summarize, 
the high speed ($v_{r}\gtrsim 0.1[c]$) outflow is driven even from the innermost region,
but its gas density is negligibly small,
leading to a very small outflow rate compared with inflow rate.

Finally, let us comment on the radial density profiles in other studies.
According to the slim disk model, mass density on the equatorial plane is expressed
as $\rho^{\rm slim}(r)\propto R^{-3/2}$ from equation (\ref{eq-slim-rho}).
Since the scale-height of the slim disk is $H \sim R$,
we expect that density at the surface is roughly proportional to
the density at the equatorial plane.
That is, mass density at the outflow launching site should rapidly grow inward in the slim disk model.
This does not agree with our simulation study, which shows
much flatter density profile.
In the GR-R-MHD simulation,
by contrast,
density profile seems flat, since we find roughly $\Sigma \propto r$
(see Figure 10 of S\c{a}dowski et al. 2015).

Much flatter density profile in our results is due probably to 
the occurrence of radial convection. This is very plausible to
occur, since entropy increases inward (in the direction of gravity),
condition for convectively unstable (see Narayan \& Yi 1994).
Note that 
convection is not taken into account in the slim-disk formulation.
It is not yet clear why convection is not so efficient in the
GR-R-MHD simulation. Careful simulation work is needed.

\subsection{The convection in super-Eddington accretion flow}

\begin{figure}[h]
\begin{center}
  \includegraphics[width=80mm,bb=0 0 1275 1275]{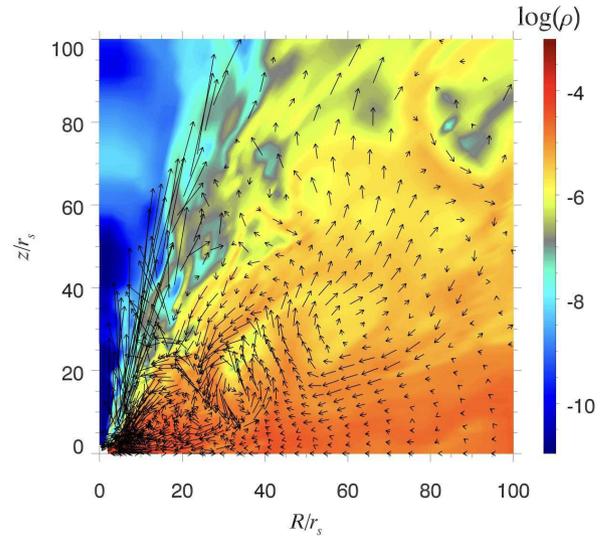}
\end{center}
\caption{
  The snapshot of the color contour map about the mass density with the arrows of the velocity in Model (a12).
  The length of the arrow corredponds to the amplitude of the velocity.
  There are circulating motions in the accretion flow.
  We remove a part of the arrows which are outside the boundaries of figure.
}
\label{fig-snap-convect}
\end{figure}

Figure \ref{fig-snap-convect} shows the convection (circulating motion) in the super-Eddington accretion disk.
We estimate the time scale of the convetion $t_{\rm conv}$ and the radiative diffusion $t_{\rm diff}$.
The convection time scale is calculated as $t_{\rm conv}=D_{\rm conv}/v\sim 0.54[{\rm s}]$.
Here, $D_{\rm conv}\sim2\pi\times15[r_{\rm S}]$ is the typical circumference of the convection,
$v\sim 5.2\times10^{8}[{\rm cm/s}]$ is the typical velocity of the convection,
at around $(R,z)\sim (70r_{\rm S},40r_{\rm S})$ in Figure \ref{fig-snap-convect}.
While, the radiative diffusion time scale is calculated as $t_{\rm diff}=3\tau_{\rm e} H /c\sim 15[{\rm s}]$.
Here, $H\sim 80[r_{\rm S}]$ is the scale height,
$\tau_{\rm e}\sim 6.3\times10^{2}$ is the optical depth in scattering,
at the radius $R\sim70[r_{\rm S}]$.
We thus understand that the convection occurs because of the relation $t_{\rm conv}\lesssim t_{\rm diff}$.

Let us explain why convection occurs when this inequality holds.
In the disk with high mass accretion rate ($\dot{M}c^{2}\geq L_{\rm Edd}$),
photon trapping effects occur as the advection of radiation entropy.
Thus, entropy increases inward;
i.e., in the direction of the gravity,
condition for convective instability (Narayan \& Yi 1994).

From here, we consider the criterion of the convection in the slim disk model
in analogy with the ADAF in Narayan \& Yi (1994).
In Watarai (2006), the pressure of the slim disk model on the equatorial plane is
\begin{eqnarray}
  p&=&7.8\times10^{15}[{\rm dyn/cm^{2}}]\nonumber\\
  &&\times f^{-1/2}\left(\frac{\alpha}{0.1}\right)^{-1}\left(\frac{M_{\rm BH}}{M_{\odot}}\right)^{-1}
  \left(\frac{\dot{M}c^{2}}{L_{\rm Edd}}\right)
  \left(\frac{r}{r_{\rm S}}\right)^{-5/2}.\label{eq-slim-pr}
\end{eqnarray}
Here, $f$ depends on $\dot{M}, r$ and, is given by from equation (24) in Watarai (2006)
\begin{eqnarray}
  f&=&f(\hat{r})=\frac{1}{2}\left(\hat{r}^{2}+2-\hat{r}\sqrt{\hat{r}^{2}+4}\right),\\
  \hat{r}&\equiv&D\frac{r}{r_{\rm S}}\frac{L_{\rm Edd}}{\dot{M}c^{2}}.
\end{eqnarray}
Here, $D$ is the numerical coefficient ($D\sim 2.18$),
and we see that $f$ approaches unity,
as $\hat{r}$ vanishes (i.e., when $\dot{M}c^{2}/L_{\rm Edd}$ is large and $r/r_{\rm S}$ is small).
Actually, $\rho$ also depends
on $f$ and $\alpha$ in such a way that $\rho \propto f^{-3/2}\alpha^{-1}$,
but we used the approximation $f\sim 1$ and set $\alpha=0.1$ in equations (\ref{eq-slim-rho}) - (\ref{eq-slim-vphi}).

From equation (13) in Narayan \& Yi (1994), in a rotating medium, the condition for a dynamical convective instability is
\begin{eqnarray}
  N^{2}_{\rm eff}&\equiv&-\frac{1}{\rho}\frac{dp}{dr}\frac{d\ln(p^{\frac{1}{\gamma}}/\rho)}{dr} + \left(\frac{v_{\phi}}{r}\right)^{2} <0 .
\end{eqnarray}
Here, $\gamma$ is the specific heat ratio.
When we use the $p, \rho, v_{\phi}$ with factor $f,\alpha$ of the slim disk model in equations (\ref{eq-slim-pr}), (\ref{eq-slim-rho}), (\ref{eq-slim-vphi}),
$N^{2}_{\rm eff}$ becomes
\begin{eqnarray}
  N^{2}_{\rm eff}&=&\frac{v_{\phi,0}^{2}}{r_{\rm S}^{2}}\left(\frac{r}{r_{\rm S}}\right)^{-3}\left[1-g(\hat{r})\right],\\
  g(\hat{r})&=&\frac{p_{0}f(\hat{r})}{\rho_{0}v_{\phi,0}^{2}}\frac{3\gamma-1}{2\gamma}\left(\frac{5}{2}-\frac{\hat{r}}{\sqrt{\hat{r}^{2}+4}}\right)\left(\frac{2\hat{r}}{\sqrt{\hat{r}^{2}+4}}-\frac{3\gamma-5}{3\gamma-1}\right).\nonumber\\
  &&\label{eq-g}
\end{eqnarray}
Here, $\rho_{0}\equiv 1.2\times10^{-5}[{\rm g~cm^{-3}}]$,
$p_{0}\equiv 7.8\times10^{15}[{\rm dyn~cm^{-2}}]$,
$v_{\phi,0}\equiv 2.1\times10^{10}[{\rm cm~s^{-1}}]$,
with $\gamma=4/3$.
We note that the $\alpha$ coefficient of the viscosity does not appear in $N_{\rm eff}^{2}$ when we consider the dependence of $\alpha$ in $\rho, p$.

The criterion of the convection instability is satisfied in the region
$g(\hat{r}) > 1$ (i.e. $N_{\rm eff}^{2} < 0$).
We obtain $\hat{r}\lesssim 1.83$ by solving the relation $g(\hat{r}) > 1$ with $\gamma=4/3$.
This range can be represented as
\begin{eqnarray}
  r \lesssim 0.84\left(\frac{\dot{M}}{L_{\rm Edd}/c^{2}}\right)r_{\rm S}.
\end{eqnarray}
This result means that the slim disk is convectively unstable
within the photon trapping radius ($r\lesssim r_{\rm trap}$).




\subsection{Future issues}
There are a number of future issues to be discussed.
Our simulations are restricted to Newtonian dynamics, but for discussing the Blandford-Znajek
(BZ) processes which are very efficient when the central black hole is rapidly spinning
we definitely need global GR-RHD simulations.
In addition, we had better solve the MHD processes in purely 3-D dimension,
since angular momentum transport by the MHD processes could be a key to examine the
existence (or absence) of large-scale circulation (or convection motion)
and thus to constrain the radial velocity and density profiles.
Such simulation studies are extremely expensive and are impossible at present.
Hence, we need careful treatments.
For example, we may solve the innermost part by the GR-R-MHD simulations to properly solve
the gas flow dynamics near the black hole and evaluate the strengths and directions of BZ flux,
and solve the outflow dynamics in a rather large simulation box by the Newtonian R-MHD simulations.
The latter is essential to discuss spectral formation of high luminosity objects, such as ULXs,
since outflow material can Compton up-scattering of the radiation from the innermost region (Kawashima et al. 2009, 2012).
Possible line emission need to be studied (see, e.g., Pinto et al. 2016),
since it could contain fruitful information from the outflow material.

Finally we need to comment on the dependence of our results
on the adopted value of the initial angular momentum (or $r_{\rm K}$).
If the quasi-steady radius increases further when we increase $r_{\rm K}$
no significant outflow is launched from every radius, in contradiction
with the powerful jets from some ULXs
(IC342 X-1 and Holmberg II X-1, Cseh et al. 2014),
and baryonic jets from an ULS (M81 ULS1, Liu et al. 2015) and SS433.
The existence of powerful outflow (or jet) is also indicated by the
observations of ULX nebulae
(e.g., Pakull \& Mirioni 2003, Gris\'e et al. 2006, Soria et al. 2010, Cseh et al. 2012).
This issue also needs further investigation, as well. 
       
\begin{ack}       
  Numerical computations were mainly carried out on Cray XC30 at Center for Computational Astrophysics, National Astronomical Observatory of Japan.
  This work is supported in part by JSPS Grant-in-Aid 
  for Scientific Research (C) (17K05383 S. M.; 15K05036 K.O.).
  This research was also supported
  by MEXT as ’Priority Issue on Post-K computer’ (Elucidation
  of the Fundamental Laws and Evolution of the
  Universe) and JICFuS.
\end{ack}       

\section*{Appendix: Derivations of Scaling Relations}
\subsection*{Black hole mass dependence, $a$}
\label{sec-index-a}
In this appendix, we describe how to calculate the index $a$ from the simulation data.
Let us, for example, compare the results of model 1 and model 2 with the same 
$\dot M$ and $r_{\rm K}$ but with different masses of $M_{1}$ and $M_{2}$.
The power-law relation, $f(M)\propto M^{a}$, leads to
\begin{equation}
  \frac{f(M_{1}; r,\theta)}{f(M_{2}; r,\theta)} = \frac{M_{1}^{a}}{M_{2}^{a}}.
\end{equation}
Here, $f$ is any physical quantities; e.g. mass density $\rho$.
We can then calculate $a(r,\theta)$ by
\begin{equation}
  a(r,\theta)\equiv\left.\left[\log\frac{f(M_{1}; r,\theta)}{f(M_{2}; r,\theta)}\right]\right/\left[\log \left(\frac{M_{1}}{M_{2}}\right)\right], \label{mdepend-a}
\end{equation}
at each grid point of $(r,\theta)$ by comparing the values of model1 and model2.
Using equation (\ref{mdepend-a}), we can calculate the mass dependence in the following way.
\begin{enumerate}
\item We adopt three values for the black hole mass; 
   $M_{\rm BH}=10, 10^{4}$, and $10^{7}$ $(M_{\odot})$.
  That is, there are ${}_{3}{\rm C}_{2}=3$ combinations of models
    for a fixed mass injection rate ($\dot{M}_{\rm input}$).
  We thus calculate there indices, $a_{i}(r,\theta)$ (${i}=1,2$, and 3) for each $\dot{M}_{\rm input}$.
\item We average the indices $a_{\rm i}(r,\theta)$ over the three combinations for each $\dot{M}_{\rm input}$; i.e.,
\begin{eqnarray}
  \average{a(r,\theta)}&\equiv&\frac{1}{3}\sum_{\rm i=1}^{3}a_{\rm i}(r,\theta). \label{eq-index-a}
\end{eqnarray}
\item The index $\average{a}$ for each $\dot{M}_{\rm input}$ 
  is calculated by averaging $\average{a(r,\theta=\pi/2)}$ over the spatial range between
  $r=5r_{\rm S}$ and $r_{\rm qss,2}$.
  The inner boundary $5r_{\rm S}$ is chosen for removing the effects of inner boundary $r=r_{\rm in}$,
  while the outer boundary $r_{\rm qss,2}\equiv 50r_{\rm S}$ ($\dot{M}_{\rm input}=300L_{\rm Edd}/c^{2}$), $100r_{\rm S}$ ($\dot{M}_{\rm input}\geq10^{3}L_{\rm Edd}/c^{2}$)
  is chosen to remove the outflow effects around $r\sim r_{\rm qss}$
  where the mass outflow rate become large value,
  in the other word, we consider that mass inflow rate is almost independent of radius in $r\lesssim r_{\rm qss,2}$.
\item %
 We have confirmed that the derived values of $\average{a}$ for each $\dot{M}_{\rm input}$ are rather insensitive to the
 $\dot{M}_{\rm input}$ values, so we simply averaged them.
\end{enumerate}

\subsection*{Accretion-rate dependence, $b$}
\label{sec-index-b}
The derivation method of the accretion-rate dependence, $b$, is the same as that
of $a$ but we replaced $M_{\rm BH}$ by $\dot M$ and $a$ by $b$.
Here, the results of the models with $\dot{M}_{\rm input}=300L_{\rm Edd}/c^{2}$ is not used.
This is because the initial angular momentum is different among other models.
The number of the combinations of models with different $\dot{M}$ is ${}_{4}{\rm C}_{2}=6$.

\subsection*{Radial dependence, $c$, and coefficients, $A_{f}$}
The index $c$ is calculated by fitting to the radial profile of each physical quantity
by a power-law relation, $f =B_{f}(r/r_{\rm S})^{c}$, with $B_{f}$ being a constant.
The spatial range of fitting is the same as before; namely, between $r=5r_{\rm S}$ and $r_{\rm qss,2}$.

The coefficients, $A_{\rho},\cdots, A_{v_{\phi}}$,
in equations (\ref{eq-dep-rho})-(\ref{eq-dep-vphi})
are calculated in the following way.
The coefficient $B_{f}$ introduced above includes the $M_{\rm BH}$ dependent part
and the $\dot{M}_{\rm BH}$ dependent part. 
To remove such dependences, we convert $B_{f}$ to $\tilde{A}_{f}$ by
\begin{equation}
 \tilde{A}_{f} \equiv B_{f}\left(\frac{M_{\rm BH}}{M_{\odot}}\right)^{-a}
                 \left(\frac{\dot{M}_{\rm BH}}{L_{\rm Edd}/c^{2}}\right)^{-b}. \label{eq-a-tilde}
\end{equation}
Here, we use the indices, $a$ and $b$, obtained in the way already mentioned above.
Then, we calculate the coefficients, $A_{\rho},\cdots, A_{v_{\phi}}$
by averaging all $\tilde{A}_{f}$ values. 
    

\end{document}